\documentclass[sigconf]{acmart}

\usepackage{array}
\usepackage{listings}
\usepackage{outlines}
\usepackage{url}
\usepackage{amsmath,amsfonts}
\usepackage{graphicx}
\usepackage{textcomp}
\usepackage{xcolor}
\usepackage{wrapfig}
\usepackage{multirow}
\usepackage{booktabs}
\usepackage{multirow}
\usepackage{float}
\usepackage{placeins}
 \usepackage{tikz}

\usepackage{framed}
\usepackage[skip=3pt]{caption}
\usepackage{subcaption}
\usepackage{graphicx,xcolor} 
\usepackage[framemethod=tikz]{mdframed}
\usepackage[linesnumbered,ruled,vlined]{algorithm2e}
\usepackage{lipsum}
\usepackage{tikz}
\usepackage{annotate-equations}
\usepackage{nicefrac}

\usepackage{wrapfig}
\usepackage{threeparttable}

\usepackage[skip=1pt plus1pt]{parskip}

\newtheorem{observation}{Observation}

\newcommand\blfootnote[1]{%
  \begingroup
  \renewcommand\thefootnote{}\footnote{#1}%
  \addtocounter{footnote}{-1}%
  \endgroup
}

\copyrightyear{2024}
\acmYear{2024}
\setcopyright{rightsretained}
\acmConference[CF '24]{21st ACM International Conference on Computing Frontiers}{May 7--9, 2024}{Ischia, Italy}
\acmBooktitle{21st ACM International Conference on Computing Frontiers (CF '24), May 7--9, 2024, Ischia, Italy}\acmDOI{10.1145/3649153.3649187}
\acmISBN{979-8-4007-0597-7/24/05}


\begin{document}

\title{Sparse MTTKRP Acceleration for Tensor Decomposition on GPU}

\author{Sasindu Wijeratne}
\affiliation{%
  \institution{University of Southern California}
  \city{Los Angeles}
  \state{California}
  \country{USA}
}
\email{kangaram@usc.edu}


\author{Rajgopal Kannan}
\affiliation{%
  \institution{DEVCOM Army Research Lab}
  \city{Los Angeles}
  \state{California}
  \country{USA}
}
\email{rajgopal.kannan.civ@army.mil}

\author{Viktor Prasanna}
\affiliation{%
  \institution{University of Southern California}
  \city{Los Angeles}
  \state{California}
  \country{USA}
}
\email{prasanna@usc.edu}

\begin{abstract}
Sparse Matricized Tensor Times Khatri-Rao Product (spMTTKRP) is the bottleneck kernel of sparse tensor decomposition. In this work, we propose a GPU-based algorithm design to address the key challenges in accelerating spMTTKRP computation, including (1) eliminating global atomic operations across GPU thread blocks, (2) avoiding the intermediate values being communicated between GPU thread blocks and GPU global memory, and (3) ensuring a balanced distribution of workloads across GPU thread blocks. Our approach also supports dynamic tensor remapping, enabling the above optimizations in all the modes of the input tensor. Our approach achieves a geometric mean speedup of 1.5$\times$, 2.0$\times$, and 21.7$\times$ in total execution time across widely used datasets compared with the state-of-the-art GPU implementations. Our work is the only GPU implementation that can support tensors with modes greater than 4 since the state-of-the-art works have implementation constraints for tensors with a large number of modes.
\end{abstract}



\maketitle

\section{Introduction}
\blfootnote{\textbf{Distribution Statement A:} Approved for public release. Distribution is unlimited.}
Tensor Decomposition (TD) provides an intuitive method for representing multidimensional data by effectively encapsulating lower-dimensional multi-aspect structures. TD is used in various domains, including network analysis~\cite{fernandes2020tensor}, machine learning~\cite{cheng2020novel, 7891546, mondelli2019connection}, and signal processing~\cite{wen2020tensor}. Within the domain of TD, Canonical Polyadic Decomposition (CPD) has emerged as a widely used approach, with the computationally intensive Matricized Tensor Times Khatri-Rao Product (MTTKRP) being the most time-consuming kernel.

Real-world tensors often exhibit irregular shapes and nonzero value distributions, which pose significant challenges when performing spMTTKRP computations on GPU. These challenges arise from irregular memory access patterns, load imbalances among a large number of GPU threads, and the synchronization overhead associated with performing atomic operations.

Recent efforts have proposed mode-agnostic tensor optimizations to address these issues by maintaining a single tensor copy, distributing spMTTKRP computations across GPU threads, and optimizing load balancing for the overall computation~\cite{li2018parti}. However, these implementations use global atomic operations, which introduce a considerable synchronization latency between streaming multiprocessors. Additionally, these approaches increase the demands on external memory since they store the intermediate computation results in GPU global memory for future use. It introduces challenges in scalability, limiting the applicability of these approaches. As the size of the tensor increases, there is a looming risk of memory explosion, further exacerbating the scalability problem. To accommodate the irregular data access patterns inherent in each tensor mode, proposed tensor formats in the literature rely on multiple copies, often called mode-specific tensor formats~\cite{8821030, 10.1145/3295500.3356216, 8048916, doi:10.1137/18M1210691} where mode-specific optimizations are used in each tensor copy. However, replicating the original tensor across different permutations of nonzero tensor elements becomes impractical as the number of modes grows. In this paper, we compare our work against state-of-the-art mode-agnostic and mode-specific implementations as discussed in Section~\ref{baselines_exp} and Section~\ref{overall_perf_exp}. 

In our prior work~\cite{10.1145/3543622.3573179}, we have introduced FLYCOO, a tensor format tailored to accelerate spMTTKRP on Field Programmable Gate Arrays (FPGAs). FLYCOO optimizes data locality across all tensor modes when accessing the input tensor and factor matrices within the FPGA external memory. Furthermore, \cite{10.1145/3543622.3573179} proposes a dynamic tensor remapping technique that is performed during execution. This strategic tensor reordering reduces inter-processor dependencies during elementwise computations. Moreover, this approach eliminates the need for multiple tensor copies corresponding to the number of tensor modes and mitigates memory explosion arising from the large number of intermediate values generated during the execution.

In this paper, we adopt and refine the FLYCOO format to create a parallel algorithm tailored for GPUs, effectively obviating the necessity for specialized hardware. We introduce GPU-specific optimizations, facilitating load-balanced computation across GPU Streaming Multiprocessors (SMs) without global atomic operations.

\newpage
The key contributions of this work are:
\begin{itemize}
\item We introduce a novel parallel algorithm to perform spMTTKRP on GPU. Our algorithm eliminates the intermediate value communication across GPU thread blocks. It achieves 2.3$\times$ higher L1-cache throughput during the execution time compared with the state-of-the-art.

\item We introduce dynamic tensor remapping on GPU to reorder the tensor during runtime, enabling mode-specific optimizations to the tensor format. These optimizations lead to 1.2$\times$ - 1.9$\times$ higher streaming multiprocessor throughput compared with the state-of-the-art.

\item We map our proposed parallel algorithm to GPU thread blocks where each thread block can concurrently execute spMTTKRP elementwise computation without global atomic operations and perform dynamic tensor remapping without atomic operations among GPU threads.

\item Our approach achieves a geometric mean speedup of 1.5$\times$ and 2.0$\times$ in total execution time compared with the baselines with mode-specific optimizations. Our work also shows a geometric mean speedup of 21.7$\times$ in execution time compared with the state-of-the-art mode-agnostic implementations.

\end{itemize}


\section{Background and Related Work}\label{background}

\subsection{Introduction to Tensors}\label{background:intro}\label{background:decomp}

A tensor is a generalization of an array to multiple dimensions. In the simplest high-dimensional case, a tensor is a three-dimensional array, which can be visualized as a data cube. For a thorough review of tensors, refer to~\cite{kolda2009tensor}. Table~\ref{table:notation} summarizes the tensor notations.

\begin{wraptable}{r}{0.33\textwidth}
\vspace{-5mm}
\caption{Notations}
\begin{center}
\begin{tabular}{|c|c|}
\hline
     \textbf{Symbol} & \textbf{Details} \\
     \hline
          $\circ$ & vector outer product \\
     $\otimes$ & Kronecker product \\
     $\odot$ & Khatri-Rao product \\
$\mathbf{A}$ & matrix \\
$\mathbf{a}$ & vector \\
     $a$ & scalar \\
     $\mathcal{X}$ & sparse tensor \\
     $\mathcal{X}_{(d)}$ & mode-$d$ matricization of $\mathcal{X}$ \\
     \hline
\end{tabular}
\label{table:notation}
\end{center}
\vspace{-5mm}
\end{wraptable}


\subsubsection{Tensor mode} 

In Tensor Decomposition, the number of dimensions of an input tensor is commonly called the number of tensor modes. For example, a vector can be seen as a mode-1 tensor. A $N$-mode, real-valued tensor is denoted by $\mathcal{X} \in \mathbb{R}^{I_0 \times \cdots \times I_{N-1}}$. This paper focuses on tensors of mode three or higher for tensor decomposition.

\subsubsection{Indices of a nonzero tensor element}~\label{index_intro}

For a 3-mode tensor, $\mathcal{X} \in \mathbb{R}^{I_0 \times I_1 \times I_2}$, a nonzero tensor element is indicated as $x = \mathcal{X}(i_0,i_1,i_2)$. Here, $i_0$, $i_1$, and $i_2$ are the positions or coordinates of $x$ in the tensor $\mathcal{X}$, which are commonly referred to as indices of the tensor element.

\subsubsection{Tensor matricization}

$\mathcal{X}_{(n)}$ denotes the mode-$n$ matricization or matrix unfolding~\cite{favier2014overview} of $\mathcal{X}$. $\mathcal{X}_{(n)}$ is defined as the matrix $\mathcal{X}_{(n)} \in \mathbb{R}^{I_n \times (I_0 \cdots I_{n-1} I_{n+1} \cdots I_{N-1})}$ where the parenthetical ordering indicates, the mode-$n$ column vectors are arranged by sweeping all the other mode indices through their ranges.

\subsubsection{Canonical Poliyedic Tensor Decomposition (CPD)}

CPD decomposes $\mathcal{X}$ into a sum of single-mode tensors (i.e., arrays), which best approximates $\mathcal{X}$. For example, given a 3-mode tensor $\mathcal{X} \in \mathbb{R}^{I_0 \times I_1 \times I_2}$, our goal is to approximate the original tensor as $\mathcal{X} \approx \sum_{r=0}^{R-1} \mathbf{a}_r \circ \mathbf{b}_r \circ \mathbf{c}_r$, where $R$ is a positive integer and $\mathbf{a}_r \in \mathbb{R}^{I_0}$, $\mathbf{b}_r \in \mathbb{R}^{I_1}$, and $\mathbf{c}_r \in \mathbb{R}^{I_2}$.



For each of the three modes, the spMTTKRP operation can be expressed as
\begin{equation} \label{eqn_spMTTKRP0}
\mathbf{\tilde{A}} =  \mathcal{X}_{(0)} ( \mathbf{{B}} \odot  \mathbf{{C}}), \text{ } 
\mathbf{\tilde{B}} = \mathcal{X}_{(1)} ( \mathbf{{C}} \odot  \mathbf{{A}}), \text{ }
\mathbf{\tilde{C}} = \mathcal{X}_{(2)}  ( \mathbf{{A}} \odot  \mathbf{{B}}) \text{ }
\end{equation}

The alternating least squares (ALS) method is used to compute CPD. In a 3-mode tensor, CPD sequentially performs the computations in Equation~\ref{eqn_spMTTKRP0}, iteratively. This can be generalized to higher mode tensors. Note that the matricization of $\mathcal{X}$ is different for each factor matrix computation. In this paper, performing MTTKRP on all the matricizations of an input tensor is called computing MTTKRP along all the modes. The outputs $\mathbf{A}$, $\mathbf{B}$, and $\mathbf{C}$ are the factor matrices that approximate $\mathcal{X}$. $\mathbf{a}_r$, $\mathbf{b}_r$, and $\mathbf{c}_r$ refers to the $r^{\text{th}}$ column of $\mathbf{A}$, $\mathbf{B}$, and $\mathbf{C}$, respectively.

In this paper, we focus on MTTKRP on sparse tensors (spMTTKRP), which means the tensor is sparse. Note however, that the factor matrices are dense.

\subsubsection{Elementwise computation}\label{background_element_computation}

The focus of this paper is to reduce the total execution time of spMTTKRP along all the modes of the tensor. Efficiently performing the elementwise computation is described below.

\begin{wrapfigure}{r}{0.30\textwidth}
\vspace{-5mm}
  \begin{center}
    \includegraphics[width=0.30\textwidth]{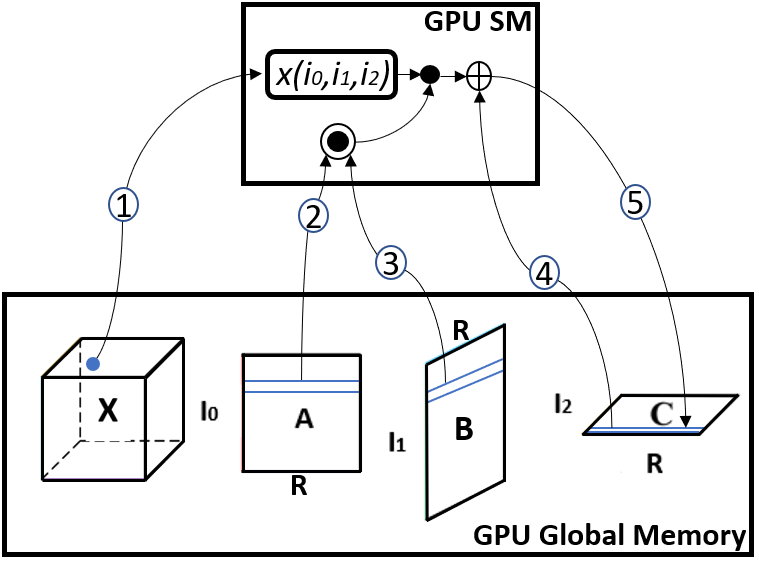}
  \end{center}
  \caption{Elementwise computation}
  \vspace{-4mm}
  \label{element_fig}
\end{wrapfigure}

Figure~\ref{element_fig} summarizes the elementwise computation of a nonzero tensor element in mode 2 of a tensor with 3 modes.

In Figure~\ref{element_fig}, the elementwise computation is carried out on a nonzero tensor element, denoted as $\mathcal{X}_{(2)}(i_0,i_1,i_2)$. In sparse tensors, $\mathcal{X}_{(2)}(i_0,i_1,i_2)$ is typically represented in formats such as COOrdinate (COO). These formats store the indices ($i_0$, $i_1$, and $i_2$) along with the element value (i.e., $val(\mathcal{X}_{(2)}(i_0,i_1,i_2))$).

To perform the computation, $\mathcal{X}_{(2)}(i_0,i_1,i_2)$ is first loaded onto the processing units (i.e., streamings multiprocessors for GPU) from the external memory (Step 1). The compute device retrieves the rows $\mathbf{{A}}(i_0,:)$, $\mathbf{{B}}(i_1,:)$, and $\mathbf{{C}}(i_2,:)$ from the factor matrices using the index values extracted from $\mathcal{X}_{(2)}(i_0,i_1,i_2)$ (Step 2, Step 3, and Step 4). Then the compute device performs the following computation:
\[
\mathbf{C}(i_2,r) = \mathbf{C}(i_2,r) + \text{val}(\mathcal{X}_{(2)}(i_0,i_1,i_2)) \cdot \mathbf{A}(i_0,r) \circ \mathbf{B}(i_1,r)
\]
Here, $r$ refers to the column index of a factor matrix row ($r < R$). The operation involves performing a Hadamard product between row $\mathbf{{A}}(i_0,:)$ and row $\mathbf{{B}}(i_1,:)$, and then multiplying each element of the resulting product by $val(\mathcal{X}_{(2)}(i_0,i_1,i_2))$. Finally, the updated value is stored in the external memory (Step 5).


\subsection{Related Work}

A. Nguyen et al.~\cite{10.1145/3524059.3532363} propose the Blocked Linearized CoOrdinate (BLCO) format that enables efficient out-of-memory computation of tensor algorithms using a unified implementation that works on a single tensor copy. In contrast to BLCO, we use a dynamic tensor format that can be used to reorder the tensor during runtime. Our work also does not require a conflict resolution algorithm like BLCO that can introduce additional overhead to the overall execution time.

I. Nisa et al.~\cite{8821030, 10.1145/3295500.3356216} propose a novel tensor format to distribute the workload among GPU threads. This work requires multiple tensor copies to perform spMTTKP along all the modes of the input tensor. Unlike~\cite{8821030, 10.1145/3295500.3356216}, our work employs a dynamic tensor remapping technique to optimize data locality during elementwise computation and eliminate the global atomic operations.

\begin{figure*}[ht]
\centering
\includegraphics[width=0.8\linewidth]{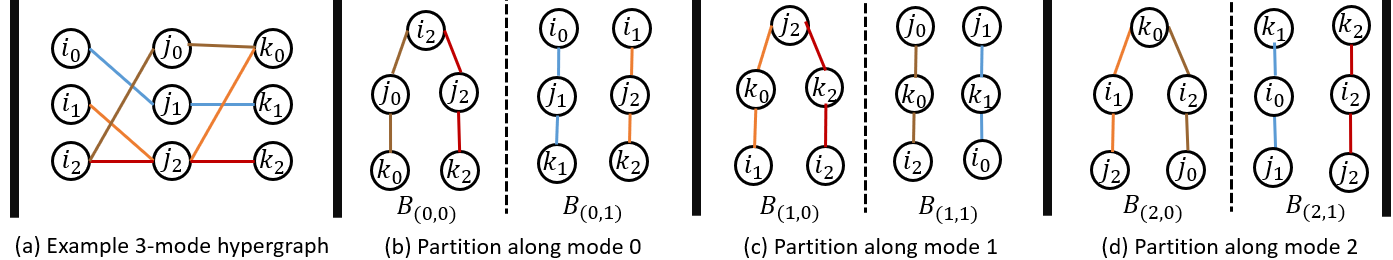}
\caption{Example hypergraph partitioning}
\label{example_hypergraph}
\vspace{-5mm}
\end{figure*}

J. Li et al.~\cite{li2018parti} introduce a GPU implementation employing HiCOO~\cite{8665782} tensor format to accelerate spMTTKRP. Their approach incorporates a block-based format with compression techniques to handle sparse tensors efficiently. Compared with~\cite{li2018parti}, our work reduces the intermediate value communication to the GPU global memory with a novel tensor format and introduces a novel tensor partitioning scheme to load balance the total computations among GPU SMs.

In our prior work~\cite{10.1145/3543622.3573179}, we developed a custom accelerator design targeted for Field Programmable Gate Array (FPGA) to perform spMTTKRP on sparse tensors. We introduce a specialized tensor format called FLYCOO, which supports custom hardware-specific optimizations. We also adopted the FLYCOO tensor format to perform spMTTKRP on multi-core CPU~\cite{wijeratne2023dynasor}. However, it is important to note that tackling spMTTKRP on a GPU presents a unique set of challenges compared to FPGA and CPU architectures. In this work, we adapt the FLYCOO format and propose GPU-specific optimizations, including ensuring a balanced distribution of workloads across GPU thread blocks, eliminating global atomic operations across GPU thread blocks, and avoiding the intermediate values being communicated across GPU thread blocks.

\section{Optimizing Tensor Format for GPU}\label{Data_partitioning}

In this paper, we develop a GPU-specific dynamic tensor remapping based on adapting the mode-agnostic tensor format FLYCOO~\cite{10.1145/3543622.3573179}. In this Section, we first introduce the dynamic tensor remapping used in FLYCOO. After that, we briefly summarize the novelty of our work following the notion of hypergraph representation of a tensor and then use it to describe our dynamic tensor remapping strategy for GPUs.

In the following, When performing spMTTKRP for mode $d$ of a tensor, we denote mode $d$ as the output mode and its corresponding factor matrix as the output factor matrix. The rest of the tensor modes are called input modes, and the corresponding factor matrices are called input factor matrices.

\subsection{Dynamic Tensor Remapping} \label{datamapping}


Dynamic tensor remapping involves reordering nonzero tensor elements at runtime based on the next mode in which spMTTKRP is performed.

Initially, the tensor is ordered based on the indices of mode 0. As the spMTTKRP computation proceeds for mode 0, the tensor is dynamically reordered according to the indices of mode 1. Consequently, when the computation for mode 1 begins, the tensor is already ordered according to the indices of mode 1.

\vspace{-4mm}
\subsection{Modified FLYCOO Tensor Format} \label{sec_novelty}
We refine the tensor element representation by introducing a novel remap ID scheme that can perform dynamic tensor remapping on each nonzero tensor element independently of each other (see Section~\ref{Tensor_Format_Representation2}). Hence, it avoids atomic operations among GPU threads while performing dynamic tensor remapping(see Observation~\ref{independent_remapping}).

We also introduce a SM-based tensor partitioning scheme that load balances the total computations among the GPU SMs (see Section~\ref{Tensor_Format_Definition}). It reduces the idle time of SMs, resulting in higher overall GPU compute throughput.
\vspace{-2mm}
\subsection{Hypergraph Representation}\label{hyper-graph} \label{sec_hypergraph}

\begin{wrapfigure}{r}{0.28\textwidth}
\vspace{-8mm}
  \begin{center}
    \includegraphics[width=0.28\textwidth]{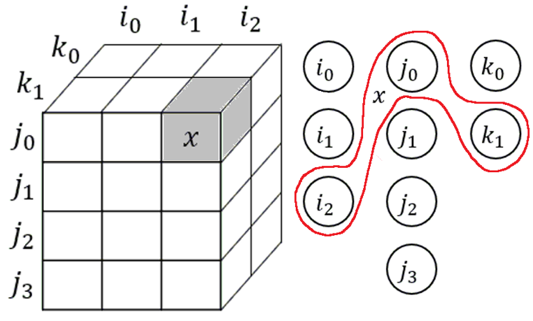}
  \end{center}
  \caption{Example hypergraph of a 3-mode tensor}
  \label{hypergraph}
  \vspace{-3mm}
\end{wrapfigure}

For a $N$ mode tensor $\mathcal{X} \in \mathbb{R}^{I_0 \times \cdots \times I_{N-1}}$, with $|\mathcal{X}|$ nonzero elements,  we consider the hypergraph, $\mathscr{G}(\textbf{I}, \Upsilon)$ with  vertex set $\textbf{I} = I_0 \cup I_1 \cup \cdots \cup I_{N-1}$ and each nonzero tensor element in $\mathcal{X}$ being represented as a hyperedge in $\Upsilon$. Here, $I_d$ is the set of all the indices in mode $d$ and  $|\Upsilon| = |\mathcal{X}|$. Figure~\ref{hypergraph} shows an example hypergraph representation of a 3-mode tensor.  


Observe that (Ref. Section~\ref{background_element_computation}) when computing spMTTKRP for a row in factor matrix of mode $d$ (the output mode), elementwise computations are performed on the nonzero tensor elements with the same mode $d$ index (Ref. Section~\ref{index_intro}) as the mode $d$ factor matrix row. In the hypergraph representation, the computation of the output factor matrix of mode $d$ involves performing elementwise operations on all the hyperedges connected to the same vertex in mode $d$ of the tensor.
Hence, we propose a partitioning scheme that brings all the hyperedges connected to the same output mode vertex into the same partition. Doing so allows each tensor partition to be executed without dependencies among tensor partitions while updating the output values.

\subsection{Tensor Partitioning Scheme} \label{Tensor_Format_Definition}

\begin{algorithm}[ht]
\DontPrintSemicolon
Input: Hypergraph $\mathscr{G}(\textbf{I}, \Upsilon)$ with vertices sorted along a given mode based on the number of hyperedges in $\Upsilon$ incident on each vertex\;
$B$ with $N \times \kappa$ empty tensor blocks\;
Output: $B$,  where each index, $i_{d,j}$ mapped to a block $B_{d,k}$\;

\For{each mode $d = 0, \ldots, N-1$} {
    \For{each vertex $j = 0, \ldots, |I_d|$} {
        \textcolor{blue}{// identify the least filled block in mode $d$}\;
        \textbf{\textit{b}} $=$ $min(|B_{d,w}|)$; $\forall w$ \;
         \textbf{\textit{b}}.append($i_{d,j}$) \;
    }
}
\Return $B$
\caption{Tensor Partitioning Scheme}
\label{ss_scheduling}
\end{algorithm}

Following the notation introduced in Section~\ref{hyper-graph}, consider the input tensor $\mathcal{X}$ and its corresponding hypergraph representation $\mathscr{G}$ where $\mathscr{G}$ is partitioned into $\kappa$ tensor partitions along each mode. In $\mathscr{G}$, for a given mode $d$, the vertices in $I_d$ are ordered based on the number of hyperedges in $\Upsilon$ incident on each vertex. Let us denote the ordered vertex set for mode $d$ as $I_{d-\text{ordered}}$. Subsequently, we iterate through the ordered list, $I_{d-\text{ordered}}$, vertex by vertex, and assign each vertex to a partition in a cyclic fashion. This step effectively partitions the vertices in mode $d$ among the $\kappa$ tensor partitions. Next, we collect all the hyperedges incident on each tensor partition. We denote these hyperedges that map to partition $j$ as Partition ID $B_{d,j}$ where $0 \leq j < \kappa$. Once the partitioning is complete, we order the hyperedges based on the partition IDs (i.e., $B_{d,j}$) and assign a \textit{remap id}, $b_d$ to each hyperedge, reflecting its position within the overall tensor. This entire process is repeated for all the modes of the hypergraph. Algorithm~\ref{ss_scheduling} summarizes the tensor partitioning scheme.

Figure~\ref{example_hypergraph} demonstrates a partitioning scheme for an example hypergraph with 4 hyperedges and 3 vertices along each mode and $\kappa = 2$. In the Figure, different hyperedges are represented by lines with different colors. In mode 0, vertex $i_2$ is incident to 2 hyperedges, while vertices $i_0$ and $i_1$ each have a single hyperedge incident to them. Following the partitioning scheme outlined in Algorithm~\ref{ss_scheduling}, we assign hyperedges incident to $i_2$ to partition 0 of mode 0 (i.e., $B_{0,0}$) and hyperedges incident to $i_0$ and $i_1$ to $B_{0,1}$. In this configuration, each partition of mode 0 contains 2 hyperedges.
This process is similarly used for the remaining modes, as shown in Figure~\ref{example_hypergraph}.

\subsubsection{Load Balancing}\label{shard_scheduling}
The proposed tensor partitioning scheme ensures a balanced load distribution among the SMs, at most 4/3 times the optimal partitioning. It also results in the same theoretical tight bound as the theorem in~\cite{graham1969bounds, 10.1145/3543622.3573179}.
\vspace{-3mm}
\subsection{Tensor Element Representation}\label{Tensor_Format_Representation2}

Using the FLYCOO tensor format in~\cite{10.1145/3543622.3573179} and  the proposed tensor partitioning scheme in Section~\ref{Tensor_Format_Definition}, a tensor $\mathcal{X}$ can be represented as a sequence $x_0, \ldots, x_{|\mathcal{X}|-1}$, where each element $x_i$ is a tuple $\langle \alpha_i$, $\beta_i$, $val_i \rangle$. Here, $\alpha_i$ $= (b_0, \ldots, b_{N-1})$ represents a vector of \textit{remap ids} based on the position of $x_i$ in each output tensor mode and $\beta_i$ $= (c_0, \ldots, c_{N-1})$ represents a vector of indices of $x_i$ in each mode (see Section~\ref{index_intro}).

\subsubsection{Memory Requirements} \label{Tensor_Format_memory}
Following the tensor element representation, a tensor element $x_i$ is a tuple $\langle \alpha_i$, $\beta_i$, $val_i \rangle$. A single nonzero element in the FLYCOO format requires $N \times \log_2(|\mathcal{X}|) + \sum_{h=0}^{N-1} \log_2 |I_h| + \delta_{\text{float}}$ bits, where $\delta_{\text{float}}$ is the number of bits needed to store the floating-point value of the nonzero tensor element.  Here, $|\alpha_i| = N \times \log_2(|\mathcal{X}|)$,  $|\beta_i| = \sum_{h=0}^{N-1} \log_2 |I_h|$, and $|val_i| = \delta_{\text{float}}$.

\section{Parallel Algorithm}\label{subsecparallel_algo}


\subsection{Elementwise Computation on GPU}
Algorithm~\ref{parallel_alg_EC} describes the elementwise computation carried out on each nonzero tensor element. In Algorithm~\ref{parallel_alg_EC}, the rows of the input factor matrices are loaded from GPU global memory (Algorithm~\ref{parallel_alg_EC}: lines 9-10) depending on the indices of the current tensor element ($\beta_i$) that is being executed in the GPU thread. Each GPU thread block locally updates the output factor matrix (Algorithm~\ref{parallel_alg_EC}: lines 15) while each thread inside the thread block maintains the coherency to ensure the correctness of the program. The elementwise computation between the tensor element and the rows of the input factor matrices (Algorithm~\ref{parallel_alg_EC}, lines 9-15) is the same as in Section~\ref{background_element_computation}.

\vspace{-3mm}
\begin{algorithm}[ht]
    \DontPrintSemicolon
\textbf{EC}($\beta_i$, $value$, $\textbf{Y}$): \;
    \textbf{Input}: Mode indices of $x_i$, $\beta_i = (c_0, \ldots, c_{N-1})$ \; \hspace{10mm} Value of $x_i$ $value$ \;\hspace{10mm} Factor matrices $\textbf{Y} = \{Y_0, Y_1,...,Y_{N-1}\}$ \; 
    \textbf{Output}: Updated $\textbf{Y}$ \;
    
    \textcolor{blue}{// $\ell$ is a vector of size \textit{R}} \;
    \For{each rank $r$ in $R$ \textbf{parallel}}{
    $\ell(r) \leftarrow value $ \;
    }
    \For{input mode $w\in\{0,\ldots,N-1\}\setminus\{d\}$}{

        $vec \leftarrow $ \textbf{Load}(row $c_w$ from $w^\text{th}$ factor matrix) \;
        \textcolor{blue}{// Row $0$ to $R-1$ of the thread block perform independent computations} \;
        \For{each rank $r$ in $R$ \textbf{parallel}}{
            $\ell(r) \leftarrow \ell(r) \times vec(r)$ \;
        }
    }

    \For{each rank $r$ in $R$ \textbf{parallel}}{
        $Y_d(c_d, r) \leftarrow \text{Threadblock}\_\text{Update}(Y_d(c_d, r) + \ell(r))$
    }

\caption{Elementwise Computation for mode $d$}
\label{parallel_alg_EC}
\end{algorithm}
\vspace{-5mm}
\subsection{Dynamic Tensor Remapping on GPU}~\label{DTR_GPU}

\vspace{-3mm}
\begin{algorithm}[ht]
    \DontPrintSemicolon
    \textbf{DR}($x_{i}$, $b_{\text{out}}$, $T_{out}$):{ \;
    \textbf{Input:} Tensor element, ($x_{i}$) \; \hspace{10mm} Next mode position of $x_i$, $p_{\text{next mode}}$ \; \hspace{10mm} Remapping tensor $T_{out}$ \;     
    \textbf{Output:} Remapped tensor, $T_{out}$ \;
     $T_{out} \leftarrow$ $x_i \cup T_{out}$ at $b_{\text{out}}$ \;
    }
\caption{Dynamic Tensor Remapping}
\label{parallel_alg_DR}
\end{algorithm}

Algorithm~\ref{parallel_alg_DR} shows executing dynamic tensor remapping on a nonzero tensor element. As described in Section~\ref{datamapping}, dynamic tensor remapping reorders the tensor during execution time to support the spMTTKRP computation along the subsequent mode. Algorithm~\ref{parallel_alg_DR} shows the dynamic tensor remapping performed during mode $d$ elementwise computation. Hence, Algorithm~\ref{parallel_alg_DR} remap the tensor according to the remap ids of mode $b_{out} = p_{(d+1) \text{ mod } N}$ to support the spMTTKRP computation along the subsequent mode $(d+1)$ mod $N$. The reordered nonzero tensor elements are collected in the tensor copy, $T_{out}$ (Algorithm~\ref{parallel_alg_DR}: line 6). With the proposed tensor partitioning scheme in Section~\ref{Tensor_Format_Definition}, all the threads in a GPU thread block that perform dynamic tensor remapping can independently operate on nonzero elements, avoiding atomic operations among the GPU threads as demonstrated in Observation~\ref{independent_remapping}.

\subsection{Parallel Algorithm Mapping to GPU Thread Blocks}~\label{sec_grid_model}~\label{sec_proposed_design}
\vspace{-5mm}
\begin{algorithm}[ht]
    \DontPrintSemicolon
\textbf{Thread Block}($B_{d,z}$, $\textbf{Y}$, $T_{out}$):{ \;
    \textbf{Input}: Input tensor partition, $B_{d,z}$\; \hspace{10mm} Factor matrices $\textbf{Y} = \{Y_0, Y_1,...,Y_{N-1}\}$\; \hspace{10mm} Remapping tensor, $T_{out}$ \;

    \textbf{Output}: Updated factor matrix of mode $d$, $Y_d$\; \hspace{12mm} Remapped tensor, $T_{out}$

$nnz \leftarrow 0$ \;
\For{$nnz < |B_{d,z}|$ \textbf{parallel}}{
    \For{each column, $t$ in thread block \textbf{parallel}}{
        \If{$nnz + t < |B_{d,z}|$}{
        \textbf{Load}($x_i$ at $(nnz+t)$) \;
            $value \leftarrow val_i$ \;
            $\beta_i = (c_0, \ldots, c_{N-1})$ \;
            $\alpha_i =$ ($b_0$, \ldots, $b_{N-1}$) \;
            $b_{\text{out}} \gets b_{(d+1)mod(N)}$ \;
            \textcolor{blue}{// Algorithm~\ref{parallel_alg_EC} \&~\ref{parallel_alg_DR} are executed in parallel} \;
            $Y_d \leftarrow$ \textbf{EC}($\beta_i$, $value$, $\textbf{Y}$) \;
            \If{thread block raw $ = R-1$}{
                $T_{out} \leftarrow$ \textbf{DR}($x_{i}$, $b_{\text{out}}$, $T_{out}$) \;
            }
    }
    }
     \textcolor{blue}{// P is the number of columns in a thread block} \;
    $nnz \leftarrow nnz + P$ \;
    }
}

\caption{Parallel Algorithm on GPU thread block (for mode $d$)}
\label{parallel_alg_grid_op}
\end{algorithm}

The basic computing unit of a GPU is a thread. According to the GPU programming model, a multi-threaded program is partitioned into blocks of threads (i.e., thread blocks) that operate independently. Thread blocks are organized into a multi-dimensional grid. For a thorough overview of the GPU programming model, please refer to~\cite{cuda2021cuda, ansorge2022programming}.

\begin{wrapfigure}{r}{0.25\textwidth}
\vspace{-5mm}
  \begin{center}
    \includegraphics[width=0.25\textwidth]{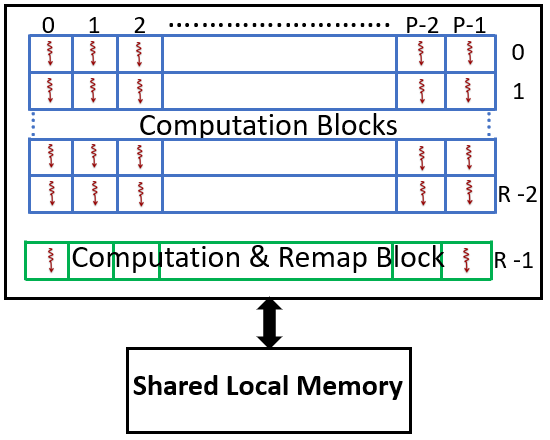}
  \end{center}
  \caption{Mapping algorithm to thread blocks}
  \label{grid_fig}
  \vspace{-5mm}
\end{wrapfigure}

We propose a GPU implementation where GPU thread blocks can perform Elementwise Computation (i.e., Algorithm~\ref{parallel_alg_EC}) and Dynamic Tensor Remapping (i.e., Algorithm~\ref{parallel_alg_DR}). Figure~\ref{grid_fig} shows a thread block with the dimensions of $R \times P$, where $R$ denotes the rank of the factor matrices and $P$ indicates the number of nonzero tensor elements parallelly loaded to a thread block. In Figure~\ref{grid_fig}, each thread corresponds to a distinct square within the thread block. Each column of the thread block shares the same nonzero tensor element. In the Figure~\ref{grid_fig}, we indicate the threads that only perform the elementwise computation in blue and the threads that perform elementwise computation with dynamic tensor remapping in green.

Algorithm~\ref{parallel_alg_grid_op} outlines the computations executed on each GPU thread block. In Algorithm~\ref{parallel_alg_grid_op}, $B_{d,z}$ corresponds to $z^{\text{th}}$ tensor partition in mode $d$ (see Section~\ref{Tensor_Format_Definition}). When a GPU SM is idle, a thread block and its corresponding tensor partition are assigned to the SM for computation. Once a tensor partition is assigned for computation, the thread block performs elementwise spMTTKRP computation and dynamic tensor remapping on the assigned partition. Each column in the thread block loads a single nonzero tensor element at a time and shares it across the threads in the same column. Each thread column extracts the embedded information from the loaded nonzero tensor element (Algorithm~\ref{parallel_alg_grid_op}: line 12-15). Subsequently, each thread block performs elementwise computation (Algorithm~\ref{parallel_alg_grid_op}: line 17). Only the last row ($R-1$) of the thread block performs dynamic tensor remapping (Algorithm~\ref{parallel_alg_grid_op}: line 18-19) on each loaded tensor element. To achieve threadwise parallelism in elementwise computation, each thread in a column only executes the computations on its corresponding rank (Algorithm~\ref{parallel_alg_EC}: line 12 - 15).

According to Algorithm~\ref{parallel_alg_grid_op}, the dynamic tensor remapping and the elementwise computation update data in the memory during the execution time. Since there are multiple thread blocks operating in parallel, the threads should not cause any race conditions while updating the data to maintain the correctness of the program. In our work, we avoid race conditions in dynamic tensor remapping and elementwise computation as follows: \\
(1) Dynamic tensor remapping: GPU threads in all the thread blocks update different locations of tensor copy $T_{out}$ during the execution time using the unique remap IDs embedded in nonzero tensor elements as discussed in Section~\ref{Tensor_Format_Definition}. It leads to avoiding atomic operations in the implementation during dynamic tensor remapping (See Observation~\ref{independent_remapping}). Atomic operations are used to prevent race conditions between threads in the same thread block or different thread blocks~\cite{nishitsuji2023basics, cook2012cuda}, which leads to synchronization overheads. \\
(2) spMTTKRP elementwise computation: During the spMTTKRP elementwise computation, multiple threads can simultaneously update the same row of the output factor matrix. Therefore, we need atomic operations among the threads, ensuring the correctness of spMTTKRP elementwise computation. Since each tensor partition is assigned to a single thread block, the proposed Algorithm~\ref{parallel_alg_grid_op} eliminates the need for global atomic operations among GPU thread blocks (See Observation~\ref{partition_independent}). Global atomic operations are used to prevent conflicts while updating values between threads in different thread blocks~\cite{nishitsuji2023basics, cook2012cuda}. Global atomic operations lead to significant synchronization overhead among threads in different GPU thread blocks.

\begin{observation}~\label{independent_remapping}
For a $N$ mode input tensor $\mathcal{X}$ in FLYCOO format, the GPU threads can perform dynamic tensor remapping for each mode $d$ ($0 \leq d < N$) without atomic operations among any GPU threads.
\vspace{-2mm}
\end{observation}
Proof: According to the FLYCOO tensor representation discussed in Section~\ref{Tensor_Format_Representation2}, we define $x_i$ as $\langle \alpha_i, \beta_i, val_i \rangle$, representing a nonzero tensor element, each $x_i$ has a distinct remap id, denoted as $b_{d}$, which denotes the location of $x_i$ in $T_{out}$ during the dynamic tensor remapping process. As per the tensor partitioning scheme defined in Section~\ref{Tensor_Format_Definition}, it is guaranteed that $b_{d}$ is a unique remap ID for $x_i$ in mode $d$. Consequently, the thread responsible for dynamic tensor remapping of $x_i$ can independently perform $x_i \cup T_{out}$ at the location $b_{d}$ without interference from other GPU threads. Given that this condition holds for all $x_i \in \mathcal{X}$, dynamic tensor remapping for tensor $\mathcal{X}$ can be executed without the need for atomic operations.

\vspace{3mm}
\begin{observation}~\label{partition_independent}
Elementwise computations of spMTTKRP can be performed without global atomic operations among GPU thread blocks.
\vspace{-4mm}
\end{observation}
Proof: Consider tensor element $x_i \in B_{d,j}$ where  $B_{d,j}$ is a partition of the input tensor $\mathcal{X}$ in mode $d$.
Let the index of $x_i$ in mode $d$ be $c_d$ where  $x_i$ update the $c_d^{\text{th}}$ row of output factor matrix of mode $d$ during the elementwise computation. Consequently, race conditions for $x_i$ can only occur with threads that execute nonzero tensor elements with index $c_d$. According to the tensor partitioning scheme described in Section~\ref{Tensor_Format_Definition}, all the tensor elements with index $c_d$ are in $B_{d,j}$. Since all the nonzero tensor elements of tensor partition $B_{d,j}$ are executed on a single thread block, race conditions corresponding to row  $c_d^{\text{th}}$ only occur inside the same thread block. Thus, there is no need for global atomic operations between GPU thread blocks while executing elementwise computation on $x_i$. Given that this condition holds for all $x_i \in \mathcal{X}$, there is no need for global atomic operations among GPU thread blocks during spMTTKRP elementwise computation.
\vspace{-3mm}
\begin{algorithm}[ht]
    \DontPrintSemicolon
    Input: Input tensor ordered according to the order of \;
    \hspace{10mm} mode 0, $T_{in}$ \;
    \hspace{10mm} Randomly initialized factor matrices, \;
    \hspace{10mm} $\textbf{Y} = \{Y_0, Y_1,...,Y_{N-1}\}$\;
    Output: Updated factor matrices $\hat{\textbf{Y}} = \{\hat{Y}_0, \hat{Y}_1,...,\hat{Y}_{N-1}\}$
    \textbf{Init}($T_{out}$) \textcolor{blue}{//Initialize tensor copy for dynamic remapping} \;
    \For{each mode $d = 0, \ldots, N-1$} {
    \For{\text{each partition of mode} $d$, $B_{d,z}$ in $T_{in}$ \textbf{parallel}}{
    \{$Y_d$, $T_{out}$\} $\leftarrow$ \textbf{Thread Block}($B_{d,z}$, $\textbf{Y}$, $T_{out}$) \;
    }
    \_\_Global Barrier\_\_ \;
   \textcolor{blue}{//Prepare tensor copies for the next mode}\;
   \{$T_{out}$, $T_{in}$\} $\leftarrow$ \textbf{Swap}($T_{in}$, $T_{out}$) \;
    }
\caption{Overall Proposed Algorithm}
\label{parallel_alg}
\end{algorithm}

\vspace{-5mm}
\subsection{Overall Algorithm}
Algorithm~\ref{parallel_alg} shows the overall parallel Algorithm for performing spMTTKRP along all the modes of an input tensor on GPU. Algorithm~\ref{parallel_alg} takes (1) $T_{in}$ which is an input tensor ordered according to $b_0$, and (2) factor matrices denoted as $\textbf{Y} = \{Y_0, Y_1,..., Y_{N-1}\}$.

As shown in Algorithm~\ref{parallel_alg}, the spMTTKRP is performed mode by mode (Algorithm~\ref{parallel_alg}: line 7). Within each mode, each thread block (Algorithm~\ref{parallel_alg}: line 9) executes a tensor partition mapped to it. At the end of all the computations of a mode, the GPU is globally synchronized before the next mode's computations to maintain the correctness of the program (Algorithm~\ref{parallel_alg}: line 10). Since we perform dynamic tensor remapping, $T_{out}$ holds a tensor copy ordered according to the next mode to be computed at the end of each mode computation. Hence, the memory pointers to each tensor copy are swapped, preparing them for the subsequent mode computation (Algorithm~\ref{parallel_alg}: line 12).

\section{Experimental Results}~\label{experiments} \vspace{-8mm}
\subsection{Experimental Setup}~\label{ex_setup}
\vspace{-5mm}
\subsubsection{Platforms}\label{sec_platform}
We conduct experiments on the NVIDIA RTX 3090, featuring the Ampere architecture. The platform has 82 Streaming Multiprocessors (SMs) and 10496 cores running at 1.4 GHz, sharing 24 GB of GDDR6X global memory. Table~\ref{table_platforms} shows the details of the platform.

\begin{wraptable}{r}{0.30\textwidth}
 \vspace{-8mm}
\caption{Platform specifications}
\begin{tabular}{|c|c|c|c|}
 \hline
 Frequency & 1695 MHz \\ 
  \hline
 Peak Performance &  35.6 TFLOPS \\
 \hline
 On-chip Memory & 6 MB L2 Cache \\
 \hline
 Memory Bandwidth & 936.2 GB/s \\
 \hline
\end{tabular}
\label{table_platforms}
\vspace{-3mm}
\end{wraptable}

We use a 2-socket AMD Ryzen Threadripper 3990X CPU with 32 physical cores (64 threads) running at 2.2 GHz, sharing 256 GB of external CPU memory for preprocessing the input tensors.

\subsubsection{Implementation}
We develop the source code using the CUDA C++~\cite{ansorge2022programming} and compile it using CUDA version 11.8~\cite{fatica2008cuda}.

\subsubsection{Datasets}

We use tensors from the Formidable Repository of Open Sparse Tensors and Tools (FROSTT) dataset~\cite{frosttdataset} and Recommender Systems and Personalization Datasets~\cite{ 10.1145/2872427.2883037, 10.1145/1376616.1376746, 10.1145/3460231.3474267, recommend_repo}. Table~\ref{table3} summarizes the characteristics of the tensors.

 \begin{table}[ht]
 \vspace{-3mm}
\caption{Characteristics of the sparse tensors}
\begin{center}
\resizebox{\columnwidth}{!}{
\begingroup
\setlength{\tabcolsep}{6pt} 
\renewcommand{\arraystretch}{1.5} 
\begin{tabular}{|c|c|c|}
 \hline
 \textbf{Tensor Name} & \textbf{Shape} & \#\textbf{NNZs} \\
 \hline\hline
 Amazon ratings only (Amazon)~\cite{10.1145/2872427.2883037, recommend_repo} & $15.2M \times 43.5M \times 7.8K$ & $233.1M$ \\
 \hline
 Delicious~\cite{frosttdataset} & $532.9K \times 17.3M \times 2.5M \times 1.4K$ & $140.1M$ \\
 \hline
 Freebase Music (Music)~\cite{10.1145/1376616.1376746} & $23.3M \times 23.3M \times 166$ & $99.5M$ \\
 \hline
 Nell1~\cite{frosttdataset} & $2.9M \times 2.1M \times 25.5M$ & $143.6M$ \\ 
 \hline
 Twitch~\cite{10.1145/3460231.3474267, recommend_repo} & $15.5M \times 6.2M \times 783.9K \times 6.1K \times 6.1K$ & $474.7M$ \\
  \hline
 Vast~\cite{frosttdataset} & $165.4K \times 11.4K \times 2 \times 100 \times 89 $ & $26M$ \\
 \hline
\end{tabular}
\endgroup
}
\label{table3}
\end{center}
\vspace{-3mm}
\end{table}

\subsubsection{Baselines}~\label{baselines_exp}
We evaluate the performance of our work by comparing it with the state-of-the-art GPU implementations: BLCO~\cite{10.1145/3524059.3532363}, MM-CSF~\cite{8821030}, and ParTI-GPU~\cite{li2018parti}. To achieve optimal results with ParTI-GPU, we use the recommended configurations provided in the source code~\cite{hicoo_repo}. For our experiments, we utilize the open-source BLCO repository~\cite{blco_github}, ParTI repository~\cite{hicoo_repo}, and MM-CSF~\cite{mm_csf} repository. The BLCO~\cite{10.1145/3524059.3532363} repository allows running MTTKRP mode-by-mode (i.e., mode-specific MTTKRP) where the input tensor is ordered specific to the given mode before running MTTKRP on GPU~\cite{blco_github}.



\subsubsection{Default Configuration}~\label{standard_config}
We use RTX 3090 with $P = 32$, $\kappa = 82$, and $R = 32$ as our configuration for conducting the experiments.


\subsection{Performance of Dynamic Remapping}~\label{ODT}
\begin{wrapfigure}{r}{0.28\textwidth}
\vspace{-3mm}
\centering
\includegraphics[width=\linewidth]{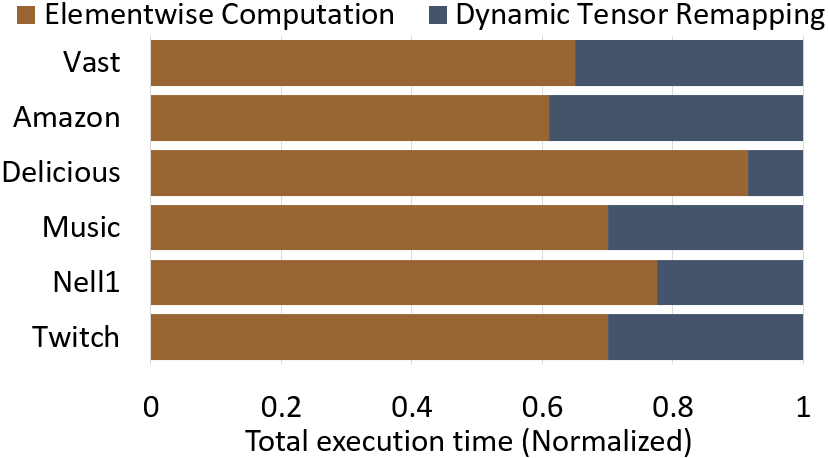}
\caption{Execution time breakdown}
\label{fig:single_copy_speedup_a}
\vspace{-5mm}
\end{wrapfigure}

Figure~\ref{fig:single_copy_speedup_a} shows a detailed breakdown of the total execution time (normalized) between elementwise computation and dynamic tensor remapping. To determine the execution time of elementwise computations in each mode, we use a mode-specific tensor copy for the computations in that mode where each tensor copy is in FLYCOO format and ordered according to the \textit{remap id} (see Section~\ref{Tensor_Format_Definition}) of the corresponding mode.

As shown in Figure~\ref{fig:single_copy_speedup_a}, the remapping overhead ranges from $5\%$ to $35\%$ for all the datasets. The overhead of dynamic tensor remapping is significantly reduced due to the thread block design (see Section~\ref{sec_grid_model}) and the tensor partitioning scheme (see Section~\ref{Tensor_Format_Definition}).

\vspace{-4mm}
\subsection{Compute SM Throughput}~\label{SMU}
\vspace{-3mm}
\begin{figure}[ht]
\vspace{-3mm}
\centering
\includegraphics[width=\linewidth]{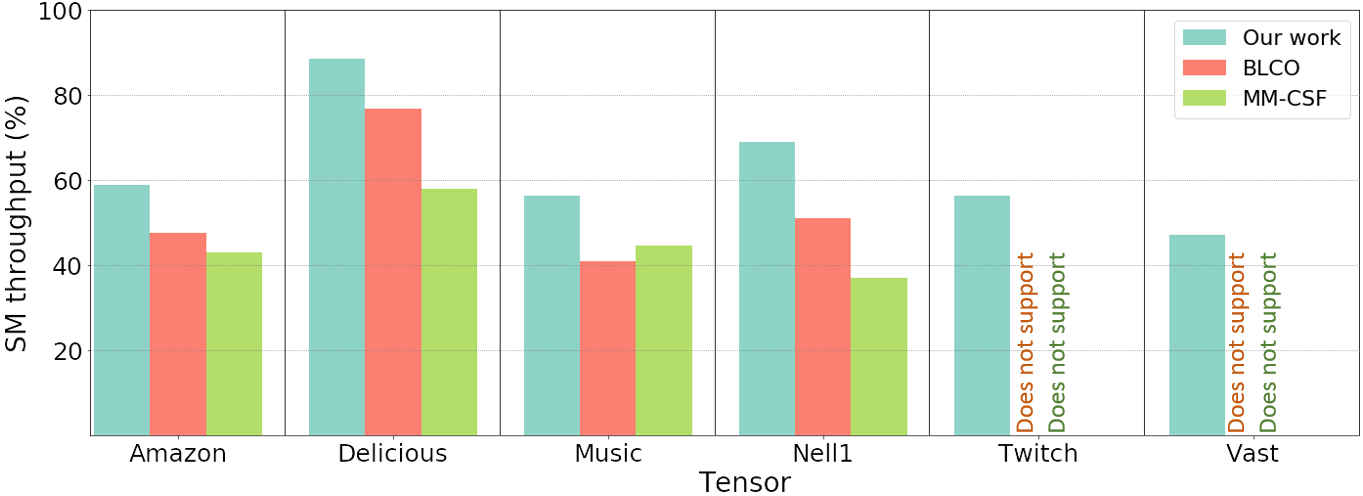}
\vspace{-5mm}
\caption{SM throughput comparison}
\label{smu_fig}
\vspace{-6mm}
\end{figure}

Compute SM throughput is a commonly used metric introduced by NVIDIA Nsight Compute~\cite{iyer2016gpu} for GPU to report the utilization achieved by the SMs while executing a kernel with respect to the theoretical maximum utilization of the selected GPU~\cite{iyer2016gpu}. NVIDIA Nsight Compute provides the achieved throughput of the kernel as a percentage value.

Figure~\ref{smu_fig} compares the SM throughput of our work for each dataset against the state-of-the-art. We use NVIDIA Nsight Compute to measure the throughput, as mentioned above. In all the datasets, our work shows 1.2$\times$ - 1.4$\times$ and 1.3$\times$ - 2.0$\times$ higher compute throughput than BLCO and MM-CSF, respectively. Our work shows higher throughout due to the minimum SM idle time of the proposed load balancing scheme and eliminating the intermediate results communication between the SMs. Since the baselines do not support tensors with a large number of modes, we could not report the SM throughput values for BLCO and MM-CSF on Twitch and Vast.

\subsection{L1 Cache Throughput}~\label{L1-util}
\begin{figure}[ht]
\vspace{-3mm}
\centering
\includegraphics[width=\linewidth]{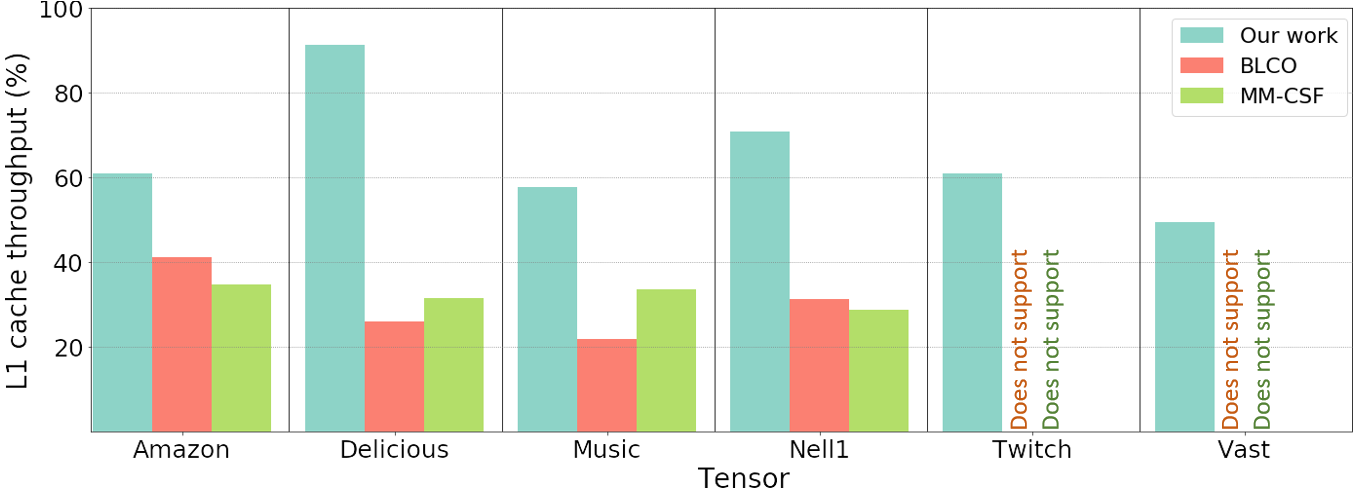}
\caption{L1 cache throughput}
\label{L1-util_fig}
\vspace{-6mm}
\end{figure}

L1 cache throughput is defined as the sustained memory throughput between all the L1 caches and their connected SMs as a percentage of the maximum theoretical throughput that can be achieved~\cite{nvidia_doc0} during the execution time of a kernel. We use NVIDIA Nsight Compute to evaluate the L1 cache throughput. Figure~\ref{L1-util_fig} shows the L1 cache throughput comparison of our work against the baselines. In all the datasets, our work shows 1.5$\times$ - 2.7$\times$ and 1.7$\times$ - 3.0$\times$ higher L1 cache throughput compared with BLCO and MM-CSF. This is due to the significant amount of data in the L1 cache being reused during the execution time.

\subsection{Impact of Algorithm to GPU Block Mapping}~\label{scalability}
\begin{figure}[ht]
\vspace{-3mm}
\centering
\includegraphics[width=\linewidth]{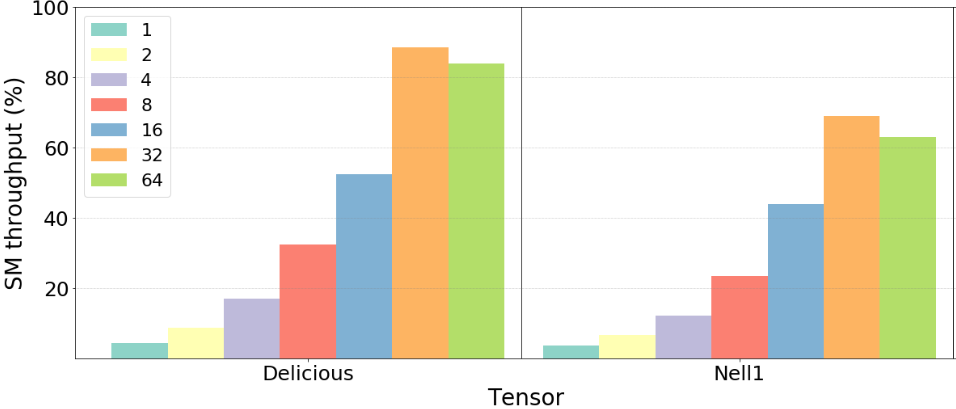}
\caption{Impact of the GPU block design}
\label{scalability_fig}
\vspace{-3mm}
\end{figure}

In our work, we map our computational model onto the GPU thread block. The number of columns in the thread block is set to match the parallel loading of nonzero tensor elements ($P$). Figure~\ref{scalability_fig} illustrates the impact of varying $P$ on SM throughput for $R = 32$. In our thread block design, R equals the number of rows in a thread block (see Section~\ref{sec_grid_model}).

We observe a linear increase in throughput as $P$ increases from $1$ to $32$. However, for $P = 64$, a decrease in throughput is noted due to multiple elementwise computations associated with different columns of the output factor matrix map into the same row of the thread block. Note that each thread block of the RTX 3090 GPU accommodates 1024 threads. Hence, keeping $P =32$ distributes the elementwise computations among the threads (i.e., $R\times P = 1024$, when $R = 32$ and $P = 32$), optimally. It is consistent across all the tensors. Therefore, we set the parameter value $P = 32$.




\begin{figure*}[ht]
\centering
\includegraphics[width=\linewidth]{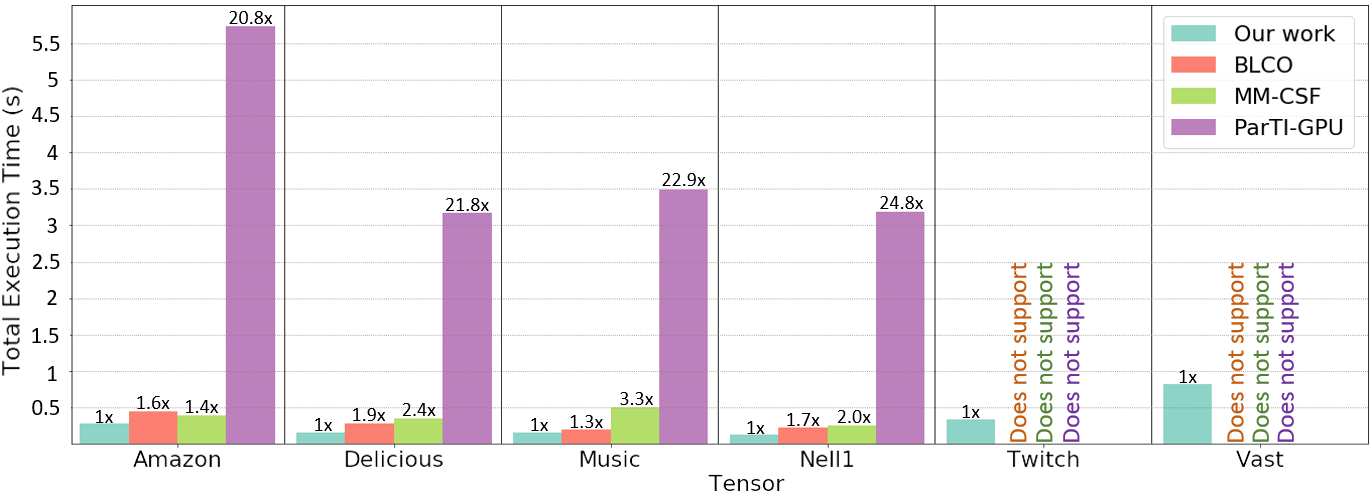}
\vspace{-4mm}
\caption{Total execution time}
\label{R32_TE}
\end{figure*}

Figure~\ref{R32_TE} shows the total execution time of our work and the baselines on the RTX 3090. The corresponding speedup achieved by our work over each baseline is displayed at the top of the respective bar. Similar to the baselines~\cite{10.1145/3524059.3532363, 8821030, li2018parti}, we set the rank of the factor matrices ($R$) to 32. Our work demonstrates a geometric mean of 1.5$\times$, 2.0$\times$, and 21.7$\times$ in speedup compared to BLCO, MM-CSF, and ParTI-GPU. Table~\ref{tab:speed-up} summarizes the overall speedup achieved by our approach compared to each baseline.

It is worth noting that MM-CSF operates as a mode-specific implementation, necessitating multiple copies of the tensor during execution. BLCO's implementation involves ordering the tensor at the beginning of each mode computation and optimizing the input tensor for efficient execution of the specific mode. These overheads of MM-CSF and BLCO are not considered in the reported timings in Figure~\ref{R32_TE}. Note that our work considers dynamic remapping overhead.

\begin{table}[ht]
\vspace{-1mm}
  \caption{Speedup of our work over state-of-the-art}
  \centering
  \begin{tabular}{|c|c|}
    \hline
     \textbf{Baseline} & \textbf{Geometric Mean} \\
    \hline
    Speedup over BLCO~\cite{10.1145/3524059.3532363} & 1.5\\
    \hline
    Speedup over MM-CSF~\cite{8821030} & 2.0\\
    \hline
    Speedup over ParTI-GPU~\cite{li2018parti} & 21.7\\
    \hline
    Overall Geometric Mean Speedup & \textbf{4.1}\\
    \hline
  \end{tabular}
  \label{tab:speed-up}
  \vspace{-5mm}
\end{table}

\subsection{Overall Performance}~\label{overall_perf_exp}
\vspace{-3mm}

Our work stands out as the only GPU implementation capable of executing large tensors with a higher number of modes, such as Twitch and Vast. In contrast, BLCO, MM-CSF, and ParTI-GPU lack support for tensors with the number of modes greater than 4.

Our work avoids communicating intermediate values among SMs and between SMs and GPU global memory. These intermediate values are stored in the L1 cache and reused with high L1 cache throughput (see Figure~\ref{L1-util_fig}). Also, our load-balancing scheme improves the overall SM throughput, reducing the idle time of the GPU SMs.

\subsection{Preprocessing Time}

The preprocessing of an input tensor involves generating tensor partitions and converting the tensor into FLYCOO format by representing a nonzero tensor element in FLYCOO tensor element representation. To accelerate this process, we use OpenMP~\cite{chandra2001parallel} and Boost library~\cite{schaling2014boost}. 

\begin{figure}[ht]
\vspace{-3mm}
\centering
\includegraphics[width=\linewidth]{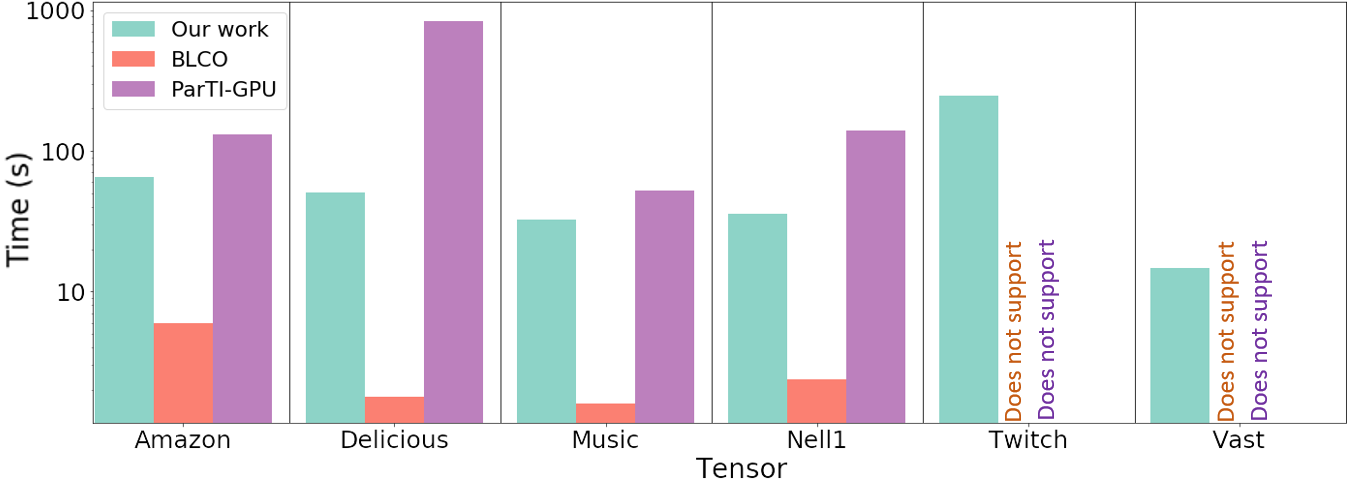}
\caption{Tensor format generation time comparison}
\label{tensor_formation}
\vspace{-5mm}
\end{figure}

Although this work does not focus on the accelerating preprocessing time, we have included a comparison of preprocessing times in Figure~\ref{tensor_formation} for the sake of completeness. For the comparison, we use the baselines that report their preprocessing time. The CPU configuration used for preprocessing can be found in Section~\ref{sec_platform}.

Our preprocessing is faster than ParTI-GPU as our preprocessing approach only looks at the nonzero tensor elements during partitioning. In contrast, the ParTI-GPU partitioning scheme~\cite{li2018parti} spans the entire index space across all the modes of a tensor, which is much larger than the number of nonzero tensor elements.

As described in Section~\ref{Tensor_Format_Definition}, we partition the tensor along all the modes. BLCO~\cite{10.1145/3524059.3532363} partitions the tensor once before executing spMTTKRP along a specific mode. Hence, BLCO preprocesses the tensor faster than our work. Note that we compare the preprocessing time of BLCO to order the tensor for a single mode.

\section{Conclusion and Future Work}~\label{conclusion}
This paper introduced a parallel algorithm design for GPUs to accelerate spMTTKRP across all the modes of an input tensor. The experimental results demonstrate that Our approach achieves a geometric mean speedup of 1.5$\times$ and 2.0$\times$ in total execution time compared with the state-of-the-art mode-specific implementations and 21.7$\times$ geometric mean speedup with the state-of-the-art mode-agnostic implementations.

Our future work focuses on adapting the proposed parallel algorithm on heterogeneous computing platforms. It will ensure that our work can be effectively applied across various hardware.

\section*{ACKNOWLEDGEMENT}
This work is supported by the National Science Foundation (NSF) under grant CNS-2009057 and in part by DEVCOM Army Research Lab under grant W911NF2220159.

\newpage
\bibliographystyle{ACM-Reference-Format}
\bibliography{chi_bib,ref}


\end{document}